\documentclass[preprint,english,letterpaper,showpacs,floatfix]{revtex4}
\usepackage[latin1]{inputenc}
\usepackage{amsmath,amssymb}
\usepackage{epsfig}

\begin{document}
\title{An Extensive Search for Overtones in Schwarzschild Black Holes}

\author{E. Abdalla}
\email{eabdalla@fma.if.usp.br}

\author{D. Giugno}
\email{dgiugno@fma.if.usp.br}

\affiliation{Instituto de F\'{\i}sica, Universidade de S\~{a}o Paulo \\
C.P. 66318, CEP 05315, S\~{a}o Paulo-SP, Brazil}


\begin{abstract}

In this paper we show that with standard methods it is possible to obtain
highly precise results for QNMs. In particular, secondary modes are
obtained by numerical integration. We compare several results making a
detailed analysis. 

\end{abstract}

\maketitle
\section{Introduction}
Wave  propagation around black holes is an active field of research (see
\cite{Regge-57}). 
The perspective of gravitational waves detection in a near future and
the great development of numerical general relativity 
have increased even further the activity on this field \cite{aguiar}. Gravitational
waves should be especially strong when emitted by black holes. The
study of the propagation of perturbations around them is, hence,
essential to provide templates for the gravitational waves
identification. Thus, activity in this field is developping quickly
\cite{many}. 

The Schwarzschild black hole is rather well known
\cite{Regge-57,Konoplya} but it is important to get reassured about the
robustness of the methods and results. Therefore, here, we embark in a
detailed study of the secondary modes by means of the subtraction of the
first modes in time domain. The method, though very simple has not been
used in the literature due to numerical errors implicit in the
method. However, we were able to use simple standard methods to show that
the results are valid implying an acuracy of seven figures for the
dominant mode and sometimes three for the secondary one.

We have used the geometric system of units, for which
$\hbar=c=G=1$. This means that the masses have dimension of length and are
measured in metres. The conversion factor from metres to kilograms is
$c^2/G$. 

\section{Fundamental Modes and First Overtones}
We begin with a series of tables on the frequencies of the QNMs for the 
Schwarzschild black holes. We have employed, throughout our discussion,
$M=1.0m$ since the black-hole frequencies scales as $M\omega=const$. For
the grid spacing, we have used $h=10^{-2}m$. In what follows, we have used
the notation $\omega^{DOM}$ for the dominant mode, $\omega^{SEC}$ for the
first overtone and, whenever applicable, $\omega^{TER}$ for the second
overtone. 

\begin{table}
\begin{tabular}{|c|c|c|c|c|}
\hline
$\ell$ & $\omega_{DOM}$ & $\delta\omega_{DOM}$ & $\omega_{SEC}$ & $\omega_{TER}$ \\
\hline
\hline
$1$ & $-$ & $-$ & $-$ & $-$ \\
\hline
$2$ & $0.483644-0.0967590i$ & $3E-7$ & $0.470-0.300i$ & $-$ \\ 
\hline
$3$ & $0.675367-0.0964997i$ & $2E-7$ & $0.667-0.288i$ & $-$ \\
\hline
$4$ & $0.867417-0.0963923i$ & $2E-7$ & $0.859-0.287i$ & $-$ \\
\hline
$5$ & $1.059614-0.0966337i$ & $6E-7$ & $1.050-0.283i$ & $-$ \\
\hline
$6$ & $1.251891-0.0963060i$ & $9E-8$ & $1.224-0.283i$ & $-$ \\
\hline
$7$ & $1.444214-0.0962866i$ & $7E-7$ & $1.433-0.282i$ & $-$ \\
\hline
$8$ & $1.636565-0.0962724i$ & $1E-6$ & $1.629-0.288i$ & $1.584-0.422i$ \\
\hline
$9$ & $1.828959-0.0962639i$ & $1E-6$ & $1.822-0.288i$ & $1.791-0.404i$ \\
\hline
$10$ & $2.021329-0.0962568i$ & $1E-6$ & $2.014-0.288i$ & $1.990-0.405i$ \\
\hline
$11$ & $2.213730-0.0962522i$ & $1E-6$ & $2.207-0.288i$ & $2.174-0.440i$ \\
\hline
$12$ & $2.406139-0.0962487i$ & $1E-6$ & $2.400-0.288i$ & $2.371-0.440i$ \\
\hline
\end{tabular}
\caption{Frequencies for the Schwarzschild BH of $M=1.0m$. Scalar Field,
  different $\ell$ values.} 
\label{TableScal1}
\end{table}

\begin{table}
\begin{tabular}{|c|c|c|c|c|}
\hline
$\ell$ & $\omega_{DOM}$ & $\delta\omega_{DOM}$ & $\omega_{SEC}$ & $\omega_{TER}$ \\
\hline
\hline
$1$ & $0.248229-0.0924905i$ & $-$ & $-$ & $-$ \\
\hline
$2$ & $0.457595-0.0950044i$ & $5E-8$ & $0.440-0.290i$ & $-$ \\ 
\hline
$3$ & $0.656899-0.0956165i$ & $2E-8$ & $0.648-0.286i$ & $-$ \\
\hline
$4$ & $0.853096-0.0958605i$ & $2E-7$ & $0.844-0.285i$ & $-$ \\
\hline
$5$ & $1.047915-0.0959821i$ & $9E-8$ & $1.039-0.282i$ & $-$ \\
\hline
$6$ & $1.242000-0.0960523i$ & $9E-8$ & $1.232-0.282i$ & $-$ \\
\hline
$7$ & $1.435647-0.0960959i$ & $7E-7$ & $1.424-0.281i$ & $-$ \\
\hline
$8$ & $1.629012-0.0961250i$ & $6E-8$ & $1.621-0.288i$ & $1.577-0.425i$ \\
\hline
$9$ & $1.822180-0.0961452i$ & $1E-6$ & $1.815-0.288i$ & $1.788-0.407i$ \\
\hline
$10$ & $2.015214-0.0961596i$ & $1E-6$ & $2.008-0.288i$ & $1.984-0.410i$ \\
\hline
$11$ & $2.208148-0.0961714i$ & $2E-6$ & $2.199-0.285i$ & $2.164-0.442i$ \\
\hline
$12$ & $2.401004-0.0961800i$ & $1E-6$ & $2.395-0.288i$ & $2.365-0.427i$ \\
\hline
\end{tabular}
\caption{Frequencies for the Schwarzschild BH of
$M=1.0m$. Electromagnetic Field, different $\ell$ values.}
\label{TableEM1}
\end{table}

\begin{table}
\begin{tabular}{|c|c|c|c|c|}
\hline
$\ell$ & $\omega_{DOM}$ & $\delta\omega_{DOM}$ & $\omega_{SEC}$ & $\omega_{TER}$ \\
\hline
\hline
$2$ & $0.37367-0.08896i$ & $6E-6$ & $0.352-0.271i$ & $-$ \\ 
\hline
$3$ & $0.599444-0.0927031i$ & $5E-9$ & $0.586-0.278i$ & $-$ \\
\hline
$4$ & $0.809180-0.0941643i$ & $4E-8$ & $0.797-0.279i$ & $-$ \\
\hline
$5$ & $1.012297-0.0948713i$ & $3E-7$ & $1.002-0.279i$ & $-$ \\
\hline
$6$ & $1.212013-0.0952667i$ & $2E-7$ & $1.203-0.286i$ & $-$ \\
\hline
$7$ & $1.409741-0.0955106i$ & $2E-7$ & $1.401-0.286i$ & $1.361-0.432i$ \\
\hline
$8$ & $1.606202-0.0956724i$ & $3E-7$ & $1.594-0.280i$ & $1.474-0.515i$ \\
\hline
$9$ & $1.801801-0.0957836i$ & $1E-6$ & $1.795-0.287i$ & $1.770-0.411i$ \\
\hline
$10$ & $1.996796-0.0958649i$ & $2E-7$ & $1.990-0.286i$ & $1.967-0.408i$ \\
\hline
$11$ & $2.191345-0.0959263i$ & $2E-7$ & $2.188-0.287i$ & $2.154-0.415i$ \\
\hline
$12$ & $2.385555-0.0959724i$ & $2E-7$ & $2.379-0.287i$ & $2.353-0.446i$ \\
\hline
\end{tabular}
\caption{Frequencies for the Schwarzschild BH of mass $M=1.0m$. Axial
  Field, different $\ell$ values.} 
\label{TableAX1}
\end{table}

A few remarks are due: the scalar $\ell=0$ field oscillated almost
nothing, so no frequencies were computed for it at all. The second
overtone wasn't clearly visible for $\ell<6$, and even so only for
$\ell=8$ and higher we could make minimally acceptable fittings to it.

\begin{figure}[h]
\begin{center}
\rotatebox{-90}{\mbox{\epsfig{file=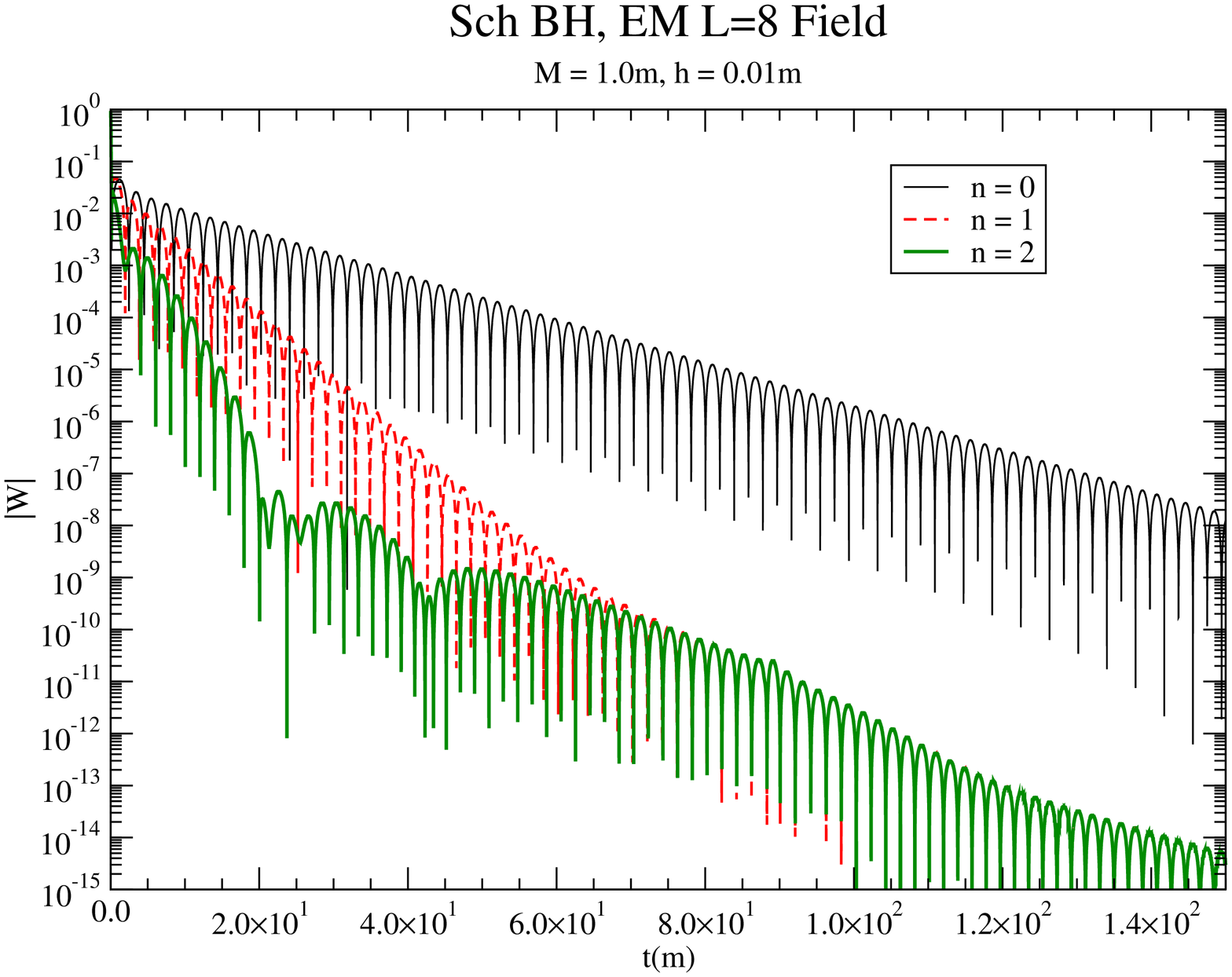,width=0.5\linewidth,clip=}}} 
\end{center}
\caption{Fundamental mode and first overtone, for the electromagnetic
  $\ell=8$ field. The second overtone is much clearer than in the previous
  figures.} 
\label{EM8}
\end{figure}

\begin{figure}[h]
\begin{center}
\rotatebox{-90}{\mbox{\epsfig{file=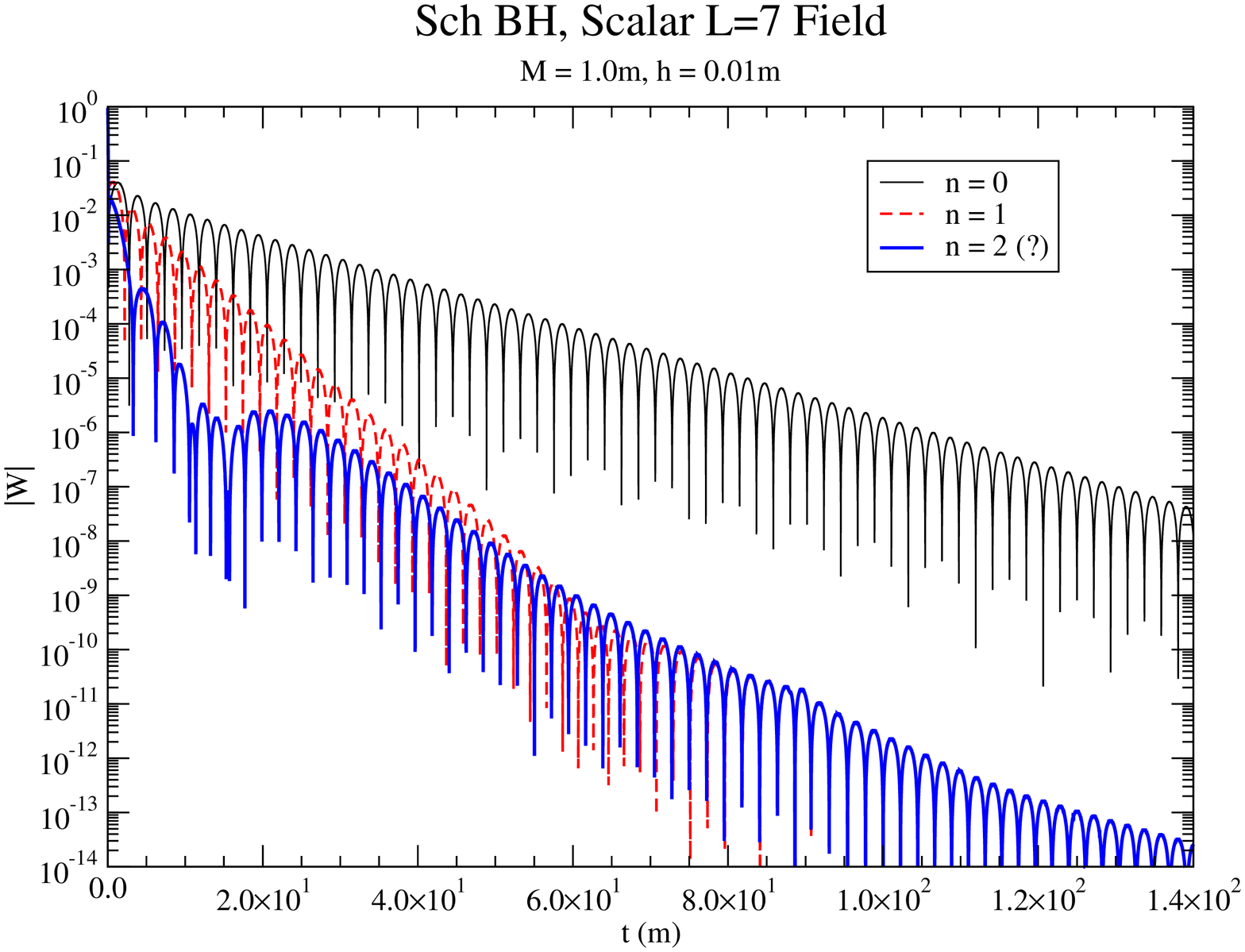,width=0.5\linewidth,clip=}}} 
\end{center}
\caption{Fundamental mode and first overtone, for the scalar $\ell=7$
  field. The second overtone also appears, though not in a very evident
  manner.} 
\label{SC7}
\end{figure}

The figures (\ref{EM8}) and (\ref{SC7}) show that the second overtone
appears clearly only for very high $\ell$ ($\ell>6$), and it is the reason
why we have blank spaces for this overtone in tables (\ref{TableScal1})
and (\ref{TableEM1}). 

We have compared our data to those generated via 6th-order WKB
computations, from \cite{Konoplya} and 3rd-order WKB from
\cite{Iyer-87}. When it comes to notation, we have employed $\ell$ for
the multipole index (as before), $n$ for the overtone index ($n=0$ for
the fundamental, $n=1$ for the first overtone, and so
on). $\omega_{NUM}$ refers to our numerical data, $\omega_{WKB1}$
refers to the WKB data from paper \cite{Konoplya} and $\omega_{WKB2}$,
to paper \cite{Iyer-87}.
 
\begin{table}
\begin{tabular}{|c|c|c|c|c|}
\hline
$\ell$ & $n$ & $\omega_{NUM}$ & $\omega_{WKB1}$ & $\omega_{WKB2}$ \\
\hline
\hline
$0$ & $0$ & $-$ & $0.1105-0.1008i$ & $0.1046-0.1152i$ \\
\hline
$1$ & $0$ & $0.292939-0.0976663i$ & $0.2929-0.0977i$ & $0.2911-0.0980i$ \\
\hline
$1$ & $1$ & $-$ & $0.264-0.307i$ & $0.2622-0.3074i$ \\
\hline
$2$ & $0$ & $0.483644-0.0967590i$ & $0.4836-0.0968i$ & $0.4832-0.0968i$ \\ 
\hline
$2$ & $1$ & $0.470-0.300i$ & $0.4638-0.2956i$ & $0.4632-0.2958i$ \\
\hline
$2$ & $2$ & $-$ & $0.4317-0.5034i$ & $0.4317-0.5034i$ \\ 
\hline
$3$ & $0$ & $0.675367-0.0964997i$ & $0.675366-0.0965006i$ & $-$ \\
\hline
$3$ & $1$ & $0.667-0.288i$ & $0.660671-0.292288i$ & $-$ \\
\hline
$4$ & $0$ & $0.867417-0.0963923i$ & $0.867416-0.0963919i$ & $-$ \\
\hline
$4$ & $1$ & $0.859-0.287i$ & $0.855808-0.290877i$ & $-$ \\
\hline
$5$ & $0$ & $1.059614-0.0963337i$ & $1.05961-0.0963368i$ & $-$ \\
\hline
$5$ & $1$ & $1.050-0.283i$ & $1.05004-0.290154i$ & $-$ \\  
\hline
$6$ & $0$ & $1.251891-0.0963060i$ & $1.25189-0.0963051i$ & $-$ \\
\hline
$6$ & $1$ & $1.224-0.283i$ & $1.24375-0.289736i$ & $-$ \\
\hline
$7$ & $0$ & $1.444214-0.0962866i$ & $1.44421-0.0962852i$ & $-$ \\
\hline
$7$ & $1$ & $1.433-0.282i$ & $1.43714-0.289473i$ & $-$ \\
\hline
$8$ & $0$ & $1.636565-0.0962724i$ & $1.63656-0.0962719i$ & $-$ \\
\hline
$8$ & $1$ & $1.629-0.288i$ & $1.63031-0.289297i$ & $$ \\
\hline
$8$ & $2$ & $1.584-0.422i$ & $1.61797-0.483757i$ & $$ \\
\hline
$9$ & $0$ & $1.828939-0.0962639i$ & $1.82893-0.0962626i$ & $-$ \\
\hline
$9$ & $1$ & $1.822-0.288i$ & $1.82333-0.289173ii$ & $$ \\
\hline
$9$ & $2$ & $1.791-0.404i$ & $1.81225-0.483235i$ & $$ \\
\hline
$10$ & $0$ & $2.021329-0.0962568i$ & $2.02132-0.0962558i$ & $-$ \\
\hline
$10$ & $1$ & $2.014-0.288i$ & $2.01625-0.289083i$ & $$ \\
\hline
$10$ & $2$ & $1.990-0.405i$ & $2.0062-0.482854i$ & $$ \\
\hline
$11$ & $0$ & $2.213730-0.0962522i$ & $2.21372-0.0962507i$ & $-$ \\
\hline
$11$ & $1$ & $2.207-0.288i$ & $2.20909-0.289015i$ & $$ \\
\hline
$11$ & $2$ & $2.174-0.417i$ & $2.19989-0.482567i$ & $$ \\
\hline
$12$ & $0$ & $2.406139-0.0962487i$ & $2.40613-0.0962467i$ & $-$ \\
\hline
$12$ & $1$ & $2.400-0.288i$ & $2.40186-0.288963i$ & $$ \\
\hline
$12$ & $2$ & $2.371-0.440i$ & $2.39338-0.482347i$ & $$ \\
\hline
\end{tabular}
\caption{Frequencies for the Schwarzschild BH of $M=1.0m$. Scalar
Field, different $\ell$ and $n$ values. Comparing to WKB data.}
\label{WKBComp1}
\end{table}

\begin{table}
\begin{tabular}{|c|c|c|c|c|}
\hline
$\ell$ & $n$ & $\omega_{NUM}$ & $\omega_{WKB1}$ & $\omega_{WKB2}$ \\
\hline
\hline
$1$ & $0$ & $0.248229-0.0924905i$ & $0.2482-0.0926i$ & $0.2459-0.0931i$ \\
\hline
$1$ & $1$ & $-$ & $0.2143-0.2941i$ & $0.2113-0.2958i$ \\
\hline
$2$ & $0$ & $0.457595-0.0950044i$ & $0.4576-0.0950i$ & $0.4571-0.0951i$ \\ 
\hline
$2$ & $1$ & $0.440-0.290i$ & $0.4365-0.2907i$ & $0.4358-0.2910i$ \\
\hline
$2$ & $2$ & $-$ & $0.4009-0.5017i$ & $0.4023-0.4959i$ \\ 
\hline
$3$ & $0$ & $0.656899-0.0956165i$ & $0.6569-0.0956i$ & $0.6567-0.0951i$ \\
\hline
$3$ & $1$ & $0.648-0.286i$ & $0.6417-0.2897i$ & $0.6415-0.2898i$ \\
\hline
$3$ & $2$ & $-$ & $0.6138-0.4921i$ & $0.6151-0.4901i$ \\
\hline
$3$ & $3$ & $-$ & $0.5814-0.6955i$ & $0.5814-0.6955i$ \\ 
\hline
$4$ & $0$ & $0.853096-0.0958605i$ & $0.853095-0.0958601i$ & $-$ \\
\hline
$4$ & $1$ & $0.844-0.285i$ & $0.841267-0.289315i$ & $-$ \\
\hline
$5$ & $0$ & $1.047915-0.0959821i$ & $1.04791-0.095981i$ & $-$ \\
\hline
$5$ & $1$ & $1.039-0.282i$ & $1.03822-0.289104i$ & $-$ \\
\hline
$6$ & $0$ & $1.242000-0.0960523i$ & $1.242-0.0960512i$ & $-$ \\
\hline
$6$ & $1$ & $1.232-0.282i$ & $1.23379-0.288982i$ & $-$ \\
\hline
$7$ & $0$ & $1.435647-0.0960959i$ & $1.43564-0.0960947i$ & $-$ \\
\hline
$7$ & $1$ & $1.424-0.281i$ & $1.42852-0.288906i$ & $-$ \\  
\hline
$8$ & $0$ & $1.629012-0.0961250i$ & $1.629-0.0961236i$ & $-$ \\
\hline
$8$ & $1$ & $1.621-0.288i$ & $1.62272-0.288855i$ & $-$ \\
\hline
$8$ & $2$ & $1.577-0.425i$ & $1.61032-0.483028i$ & $-$ \\
\hline
$9$ & $0$ & $1.822180-0.0961452i$ & $1.82217-0.0961439i$ & $-$ \\
\hline
$9$ & $1$ & $1.815-0.288i$ & $1.81655-0.288819i$ & $-$ \\
\hline
$9$ & $2$ & $1.788-0.407i$ & $1.80542-0.48265i$ & $-$ \\
\hline
$10$ & $0$ & $2.015214-0.0961596i$ & $2.01521-0.0961587i$ & $-$ \\
\hline
$10$ & $1$ & $2.006-0.288i$ & $2.01012-0.288793i$ & $-$ \\
\hline
$10$ & $2$ & $1.984-0.410i$ & $2.00003-0.482374i$ & $-$ \\
\hline
$11$ & $0$ & $2.208148-0.0961714i$ & $2.20814-0.0961697i$ & $-$ \\
\hline
$11$ & $1$ & $2.199-0.285i$ & $2.20349-0.288773i$ & $-$ \\
\hline
$11$ & $2$ & $2.161-0.442i$ & $2.19426-0.482167i$ & $-$ \\
\hline
$12$ & $0$ & $2.401004-0.096180i$ & $2.40099-0.0961782i$ & $-$ \\
\hline
$12$ & $1$ & $2.395-0.288i$ & $2.39672-0.288758i$ & $-$ \\
\hline
$12$ & $2$ & $2.365-0.427i$ & $2.38822-0.482007i$ & $-$ \\
\hline
\end{tabular}
\caption{Frequencies for the Schwarzschild BH of
$M=1.0m$. Electromagnetic field, different $\ell$ and $n$
values. Comparing to WKB values.}
\label{WKBComp2}
\end{table}

\begin{table}
\begin{tabular}{|c|c|c|c|c|}
\hline
$\ell$ & $n$ & $\omega_{NUM}$ & $\omega_{WKB1}$ & $\omega_{WKB2}$ \\
\hline
\hline
$2$ & $0$ & $0.37367-0.08896i$ & $0.373691-0.088891i$ & $0.3732-0.0892i$ \\ 
\hline
$2$ & $1$ & $0.352-0.272i$ & $0.346297-0.27348i$ & $0.3460-0.2749i$ \\
\hline
$2$ & $2$ & $-$ & $0.2985-0.4776i$ & $0.29852-0.47756i$ \\
\hline
$3$ & $0$ & $0.599444-0.0927031i$ & $0.599443-0.0927025i$ & $0.5993-0.0927i$ \\
\hline
$3$ & $1$ & $0.587-0.278i$ & $0.582642-0.28129i$ & $0.5824-0.2814i$ \\
\hline
$3$ & $2$ & $-$ & $0.551594-0.479047i$ & $0.5532-0.4767i$ \\
\hline
$4$ & $0$ & $0.809180-0.0941643i$ & $0.809178-0.0941641i$ & $0.8091-0.0942i$ \\
\hline
$4$ & $1$ & $0.797-0.279i$ & $0.796631-0.284334i$ & $0.7965-0.2844i$ \\
\hline
$4$ & $2$ & $-$ & $0.7727-0.4799i$ & $0.772695-0.4799i$ \\
\hline
$5$ & $0$ & $1.012297-0.0948713i$ & $1.0123-0.0948706i$ & $-$ \\
\hline
$5$ & $1$ & $1.002-0.279i$ & $1.00222-0.285817i$ & $-$ \\
\hline
$6$ & $0$ & $1.212013-0.0952667i$ & $1.21201-0.0952659i$ & $-$ \\
\hline
$6$ & $1$ & $1.203-0.286i$ & $1.20357-0.28665i$ & $-$ \\
\hline
$7$ & $0$ & $1.409741-0.0955106i$ & $1.40974-0.0955096i$ & $-$ \\
\hline
$7$ & $1$ & $1.401-0.286i$ & $1.40247-0.287164i$ & $-$ \\
\hline
$7$ & $2$ & $1.361-0.432i$ & $1.38818-0.480709i$ & $-$ \\
\hline
$8$ & $0$ & $1.606202-0.0956724i$ & $1.60619-0.0956707i$ & $-$ \\
\hline
$8$ & $1$ & $1.594-0.280i$ & $1.59981-0.287504i$ & $-$ \\
\hline
$8$ & $2$ & $1.474-0.515i$ & $1.58721-0.480804i$ & $-$ \\ 
\hline
$9$ & $0$ & $1.801801-0.095784i$ & $1.80179-0.0957828i$ & $-$ \\
\hline
$9$ & $1$ & $1.795-0.288i$ & $1.7961-0.287741i$ & $-$ \\
\hline
$9$ & $2$ & $1.770-0.440i$ & $1.78483-0.48087i$ & $-$ \\
\hline
$10$ & $0$ & $1.996796-0.0958649i$ & $1.99679-0.0958639i$ & $-$ \\
\hline
$10$ & $1$ & $1.990-0.286i$ & $1.99165-0.287912i$ & $-$ \\
\hline
$10$ & $2$ & $1.967-0.408i$ & $1.98145-0.48087i$ & $-$ \\
\hline
$11$ & $0$ & $2.191345-0.0959263i$ & $2.19133-0.0959245i$ & $-$ \\
\hline
$11$ & $1$ & $2.185-0.287i$ & $2.18665-0.28804i$ & $-$ \\
\hline
$11$ & $2$ & $2.154-0.415i$ & $2.17734-0.480952i$ & $-$ \\
\hline
$12$ & $0$ & $2.385555-0.0959724i$ & $2.38554-0.095971i$ & $-$ \\
\hline
$12$ & $1$ & $2.379-0.287i$ & $2.38124-0.288138i$ & $-$ \\
\hline
$12$ & $2$ & $2.353-0.446i$ & $2.37268-0.480979i$ & $-$ \\
\hline
\end{tabular}
\caption{Frequencies for the Schwarzschild BH of mass $M=1.0m$. Axial
Field, different $\ell$ and $n$ values. Comparing data to WKB.}
\label{WKBComp3}
\end{table}

These comparison tables leave no room for doubts when it comes to the
fundamental mode, $n=0$. For all fields and $\ell$-values, the data from
our numerical simulations and those from the 6th-order WKB method showed a
high-degree agreement, with differences between 1 part in $10^4$ and 1
part in $10^5$, for both $Re(\omega)$ and $Im(\omega)$, especially for
higher $\ell$-values. When it comes to the first overtone, for all fields
under study, the agreement between both sets of data was not so
impressive, hovering around $1-2\%$ for $Re(\omega)$ and $2-3\%$ for
$Im(\omega)$ for low $\ell$ (typically up to $\ell=6$) and improving for
higher $\ell$, to a few parts in 1000 for $Re(\omega)$ and around $0.5\%$
or so for $Im(\omega)$. The second overtone showed much higher
discrepancies, especially for $Im(\omega)$, with differences around $15\%$
is some cases, while for $Re(\omega)$ this difference usually hovers
around $1-2\%$ (with two exceptions for $\ell=8$, when it was much bigger
- almost $8\%$ for the axial field).

\begin{appendix}

\section{Finding the Overtones Numerically}

A comment is due on the method of extracting the secondary (and, whenever
applicable, the tertiary) mode: we took the first (or dominant) mode
($n=0$) and applied an oscillatory fitting to it, in order to extract the
data on its oscillatory frequency $\omega_{R}$, amplitude $A$ and decay
rate $\omega_{I}$. These data were taken with the maximum number of
significant digits possible (usually 8 or 9), and we subtracted the fitted
function, which had the form 
\begin{equation}
\Psi_{0}=A\exp(i\omega_{R}^{0}t)\exp(-\omega_{I}^{0}t),
\label{fit}
\end{equation}
so that we could see the remainder of this operation. In all cases (except
for the $\ell=0,1$ scalar and $\ell=1$ EM field), we could find a
remainder of the form 
\begin{equation}
\Psi_{R}=B\exp(i\omega_{R}^{1}t)\exp(-\omega_{I}^{1}t)+\Delta,
\label{2ndmode}
\end{equation}
in which $\omega_{1}\neq\omega_{0}$, characterizing a secondary mode (the
first overtone), and with similar amplitudes ($A\thickapprox B$). The term
$\Delta$ is just the $\Psi_{0}$ term, but with a much reduced amplitude,
indicating that the fitting operation has its precision limitations. This
can be seen in the form of parallel curve envelopes in the figures from
above. 

A similar procedure was adopted to find the second overtone, this time
applying the oscillatory fitting to the first overtone, in which a new
remainder similar to that seen in (\ref{2ndmode}) was seen. In this latter
case, however, only the very high $\ell$-values could yield any valuable
result when an oscillatory fitting was applied to it. 

Now it is the time to estimate the precision of the first fitting. As we
can see again in the aforementioned figures, there is a given time $t_{c}$
for which the secondary mode and the dominant mode are roughly equal in
magnitude, that is
\begin{equation}
B\exp(i\omega_{R}^{1}t)\exp(-\omega_{I}^{1}t)\thickapprox
\delta(A\exp(i\omega_{R}^{0}t)\exp(-\omega_{I}t)).  
\end{equation}

The variation of the dominant mode can be expressed as
\begin{eqnarray}
\delta(A\exp(i\omega_{R}^{0}t)\exp(-\omega_{I}t))=(\delta
A)\exp(i\omega_{R}^{0}t)\exp(-\omega_{I}t)&+&\nonumber \\ 
&+&Ai\exp(i\omega_{R}t)\exp(-\omega_{I}t)[\delta\omega_{R}^{0}+i\delta\omega_{I}^{0}]t.
\end{eqnarray} 
In what follows, we assume $\delta A$ to be much smaller than $A$. Such approximation implies
\begin{equation}
B\exp(i\omega_{R}^{1}t)\exp(-\omega_{I}^{1}t)\thickapprox
Ait\delta\omega\exp(i\omega_{R}^{0})\exp(-\omega_{I}^{1}t), 
\end{equation}
in which we have used
$\delta\omega=\delta\omega_{R}^{0}+i\delta\omega_{I}^{0}$.Working with the
moduli of the quantities above, we arrive at 
\begin{equation}
|\delta\omega|\thickapprox|\frac{B}{A}|\frac{\exp(-[\omega_{I}^{1}-\omega_{I}^{0}])t}{t}.
\end{equation}
Upon substituting $t=t_{c}$ and noting that $|A|\thickapprox|B|$, we arrive at
\begin{equation}
|\delta\omega|\thickapprox\frac{\exp(-[\omega_{I}^{1}-\omega_{I}^{0}])t_{c}}{t_{c}}.
\end{equation}
Such an estimate was done for each field and for each $\ell$-value,
and $|\delta\omega|$ was compared to $|\omega|$, so as to determine
the approximate degree of precision in the computation of the dominant
mode frequency. A similar estimate can also be done, in principle, for
the first overtone fitting precision. We have determined the ratio
$m=|\frac{\delta\omega}{\omega}|$ - for the $n=0$ mode - to be smaller than $10^{-6}$, in all
cases we have dealt with, except for the axial $\ell=2$ case, for
which $m\thickapprox 10^{-5}$ (hence its smaller number of significant figures).
\end{appendix}

\end{document}